\documentstyle[emulateapj,psfig,onecolfloat]{article}

\def\etal{{\rm et~al.\ }}
\def\hmpc{\;h^{-1}{\rm Mpc}}

\def\kms{{\rm \;km\;s^{-1}}}

\def\msun{{\rm h^{-1} M_{\odot}}}

\def\simlt{\lower.5ex\hbox{$\; \buildrel < \over \sim \;$}}
\def\simgt{\lower.5ex\hbox{$\; \buildrel > \over \sim \;$}}

\def\vg{$v_{g}$ }

\begin{document}

\twocolumn[

\title{
Gravitational Redshifts in Simulated Galaxy Clusters
}

\author{
Young-Rae Kim$^{1,3}$ and
Rupert A.C. Croft$^{1,2}$,
}

\begin{abstract}
We predict the amplitude  of the gravitational redshift of galaxies in
galaxy clusters
using an N-body simulation of  $\rm{\Lambda}$CDM
universe. We examine if it might be possible to
detect    the   gravitational   effect   on   the    total   redshift
observed for galaxies. 
 For  clusters  of mass  $M  \sim10^{15}\msun$,  the
difference in  gravitational redshift  between the brightest  galaxy and
the rest of the cluster members is $\sim10 \kms$. 
The most efficient way to detect gravitational
redshifts using information from galaxies only
involves using the full gravitational redshift
profile of clusters. 
Massive clusters, while having the largest gravitational
redshift suffer from large galaxy peculiar velocities and substructure,
which act as
a source of noise. This and their low number density make 
it more reasonable to try averaging over
many clusters and groups of relatively low mass.
We examine publicly available data for 107 rich clusters from the 
ESO Nearby Abell Clusters Survey (ENACS), finding no evidence for
gravitational redshifts.
Test on our simulated clusters
show that we need at least $\sim 2500$ clusters/groups with
$M> 5 \times 10^{13}\msun$ for a detection of
gravitational redshifts at the 2$\sigma$ level.
\end{abstract}
 
]

\footnotetext[1]{Dept.   of  Physics,   Carnegie   Mellon  University,
Pittsburgh, PA 15213} 
\footnotetext[2]{Max-Planck Institute for Astrophysics, Garching,
 Munich, D-85740, Germany} \footnotetext[3]{yr@cmu.edu}

\section{Introduction}
General  Relativity  predicts  redshift  of  photons due to a 
gravitational  field.  When  a  photon with  wavelength  $\lambda$  is
emitted in a gravitational potential  $\Phi$, it will lose energy when
it  climbs up in the  gravitational  field and  will consequently  be
redshifted. The redshift observed at infinity is given in the weak field
limit by:
\begin{equation}
z_{g} = \frac{\Delta \lambda}{\lambda} \simeq \frac{\Delta \Phi}{c^{2}}
\label{eqn:zg}
\end{equation}
where $\Delta \lambda$, $\Delta \Phi$ are respectively
the difference in wavelength,
and difference in potential between where the 
photon is emitted and where it is observed.

If we consider galaxies as sources of the photons,
the gravitational redshift effect is so tiny that we take it for granted
that  a measurement of the total galaxy redshift can be assumed to be
the sum of Hubble expansion and peculiar
velocities. In this paper we examine whether this is always the case,
and in particular whether galaxies in galaxy clusters could have 
measurable values of $z_{g}$.

 Since the gravitational  potential depends on the mass
distribution around  galaxies,  the gravitational redshift, if observable,
 should be most
evident in dense environments. 
In an early study by Nottale (1976), the redshift 
difference between pairs of clusters
was compared to the richness difference. A supposed strong
effect was found, with the pair member of
higher richness having a systematically high redshift. However, 
when Rood \& Struble (1982) rexamined this with a larger sample, 
their result showed no such correlation.
Nottale (1990)  discussed that the
effect should be looked for in galaxies at the centers of  galaxy
clusters, by comparing their redshifts with those of 
galaxies at the cluster edges. Stiavelli \& Setti (1993) 
carried out a related test in individual elliptical
galaxies, finding at $99.9\%$ confidence that 
elliptical galaxy cores are redshifted with 
respect to the galaxy outer regions, explaining this as a result 
of gravitational redshift. 

The study of gravitational redshifts in galaxy clusters was taken further
by Cappi (1995), who modelled clusters  using different
density profiles including a de Vaucouleurs law. It was predicted
that the gravitational redshift is non-negligible in very rich
clusters. For example, the centers of
clusters of masses $10^{16} \msun$ should be redshifted by
as much as $300 \kms$ with respect to infinity. 
Broadhurst \&  Scannapieco (2000) modelled the effect using
a Navarro Frenk \& White (1997) (hereafter NFW) 
density profile, and suggested that the
gravitational redshift of metal lines in the cluster gas could 
eventually be used to map out the potential directly.
As the gravitational redshift  is
sensitive to the distribution of mass in the innermost
regions of clusters, it could be  used as a  probe to
constrain the amount of dark matter there.
Gravitational redshifts would provide 
complimentary information to gravitational lensing (e.g., Sand \etal 2003),
 as unlike lensing
they do not depend on the mass density projected along the line
of sight.

Here we use an $N$-body simulation made publically 
available by the Virgo Consortium (Frenk et al.  2000) to estimate the
magnitude of the effect of gravitational redshifts
on galaxy clusters in a $\rm{\Lambda CDM}$ universe. We examine 
possible observational strategies and
determine if galaxy gravitational redshifts could be detected 
with a reasonable   number  of   clusters.  By  using   the  numerical
simulation, we will be able to study the effect of
substructure in the density
and the  potentially complex
velocity  field of  realistic clusters. We will see
if Nottale's suggestion of measuring the difference in gravitational redshift
between the  central galaxy  and galaxies at  the edge of  the cluster
is realizable in practice.
The potential wells should be deeper for the most massive clusters,
which means larger gravitational redshifts. 
However, massive clusters are rare, so that there may not
be enough large mass clusters in reality, something which we
will investigate.

The  gravitational redshift  of the galaxy closest to the
center of the potential is
always positive with respect to the 
other galaxies, whereas the Hubble velocity component
 and peculiar velocity components can have either sign.
We  extract the gravitational redshift information  by summing the
total  redshift from galaxies in many  clusters  so that  the  noise 
from  Hubble
expansion and peculiar velocities is averaged out. Small mass clusters
may be better candidates than massive ones
in  this sense because they are more abundant
in  the Universe.  The  effective  mass range  for  detection is  also
discussed in this paper.

The  structure of  the  paper is  as  follows. In  \S2, the  numerical
simulation used as our model universe is described. In \S3, we discuss
the magnitude of the gravitational  redshift and the number clusters needed
in order to detect the  effect observationally. In \S4, we briefly
examine constraints on gravitational redshifts from
an observed sample of cluster galaxies, take from the
ENACS survey. Finally,  in \S5, we summarize our results and conclude.

\section{Simulations}
\subsection{Model Universe}
Our model universe  is an output of an $N$-body simulation 
run by the Virgo Consortium(Frenk  et al. 2000), in
which  discrete  particles   represent  dark  matter.  The  background
cosmology is  dominated by a cosmological  constant
 ($\Omega_{\Lambda}=0.7$) and the  matter  content is  
85\%  Cold Dark  Matter  (CDM). No gas dynamics is included
and dissipationless particles are used to model the baryons also.
 The output we use is of  the $\Lambda$CDM  model at $z=0$
 ($\Omega_{M}=0.3,  h =
0.7,  \sigma_{8} =  0.9$). The  box  is periodic, of
side length  is 239.5$\hmpc$  and  contains
$256^{3}$  particles of mass 6.86  $\times
10^{10} \;h^{-1} \msun$.

\begin{figure}[t]
\centerline{
\psfig{file=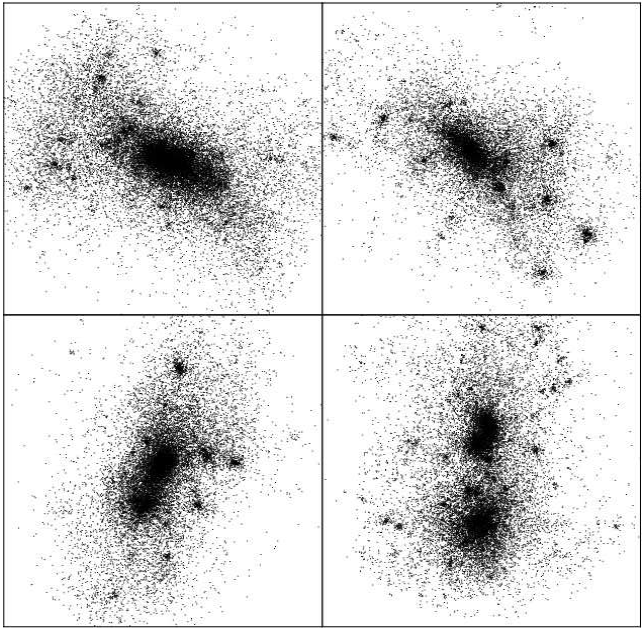,angle=-90.,width=8.5truecm}
}
\caption{ Some cluster examples. The  top left panel shows the most massive
cluster in the simulation ($M = 1.4 \times 10^{15} \msun$); top right:
$M = 8.3  \times 10^{14} \msun$; bottom left: $M  = 1.1 \times 10^{15}
\msun$; bottom  right: $M = 1.2  \times 10^{15} \msun$.  Each panel is
$8\hmpc$ wide.
\label{fig:clusterxy}
}
\end{figure}

\subsection{Cluster and Galaxy Selection}
For group finding, we use the friends-of-friends (fof) method (e.g., Huchra
\& Geller 1982). A  particle is defined to belong to a  group if it is
within  some  linking  length  ($b$)  of any  other  particle  in  the
group. We select clusters  by using  $b=0.2$ where $b$ is
the linking length as a  fraction of the mean particle separation.
This definition of dark matter haloes was shown by Jenkins \etal (2001)
to yield a mass function with a univeral form.
Running the fof algorithm again, this time with $b=0.05$,
we find the  largest group in  each cluster, and define this
to be the most massive or central  galaxy. When  we refer to ``galaxies'' 
apart from the most massive galaxy later in the paper, we refer
to    particles   randomly   sampled    from   those    cluster   member
particles which do not fall in the central galaxy halo.

In  Figure~\ref{fig:clusterxy} we show some examples of
rare rich clusters selected from the simulation in this way. The
most massive   cluster    in    the   entire simulation   has mass
$\sim1.4\times10^{15}\msun$.  Because the  simulation includes non-linear
clustering and clusters form from the bottom up,
 we expect each  cluster to have its  own substructure. Even though we are 
at $z=0$ and one might have expected some relaxation to have 
occurred, the substructure is very obvious. The cluster in the bottom
right plainly consists of two clusters with similar masses 
merging. If such a merger was seen face on, then observationally
it would be possible to separate the clusters. However, if the 
clusters lie one behind the other along the line of sight, this would
not be possible. In analyzing our simulated clusters, we are conservative and 
do not assume that clusters with much substructure could be removed, so that
we use all clusters selected with our fof algorithm. 

\begin{figure}[t]
\centerline{
\psfig{file=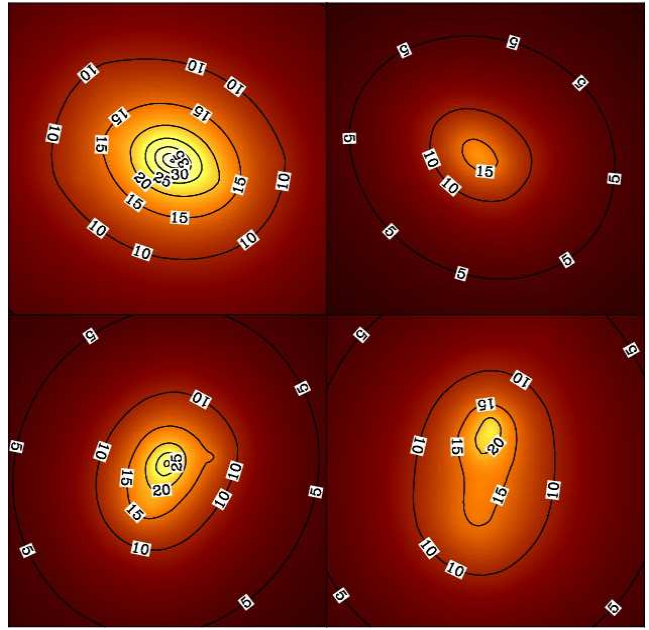,angle=-90.,width=8.5truecm}
}
\caption{ 
Slices through the
gravitational potential profile of different 
mass clusters (same as
Figure~\ref{fig:clusterxy}). Contours are of gravitational
redshift with respect to $\infty$ and are labelled in km/s.
\label{fig:pot}
}
\end{figure}

\subsection{Calculation of Gravitational Redshifts}
As a measure of gravitational redshift, we adopt 
velocity units, so that the quantity is given by:
\begin{equation}
\Delta v_{g} = \Delta z_{g} c = \frac{\Delta \Phi}{c} ,
\label{eqn:vg}
\end{equation}
where  $\Phi$  is the  gravitational  potential. 
For each particle in the simulation, we calculate the 
gravitational potential with respect  to infinity
using a tree  algorithm (Barnes  \&  Hut  1986).
When doing this, we soften the gravitational potential in
order to avoid noise from particles very close to one another,
using 5 h$^{-1}$ Kpc equivalent Plummer softening (e.g., Aarseth 1963).

In  Figure~\ref{fig:pot} we plot contours of gravitational
redshift in some examples of galaxy clusters (the same
clusters whose density distributions were shown in Figure
 ~\ref{fig:clusterxy}.)
Each panel shows the two dimensional slice of gravitational potential  in
the $xy$ plane that passes through the cluster center of mass.
The substructure in the cluster density field is also somewhat
reflected in the potential profile. However, the potentials
are smoother, and even the merging clusters in the bottom right panel
do not show strong evidence of two minima in the potential.
The gravitational redshifts, which are related to the potential 
will therefore vary smoothly across the cluster, without much substructure.
The substructure in the density distribution will however cause
noise in the Hubble flow part of the total redshift (see \S 3.2).

\begin{figure}[b]
\centerline{
\psfig{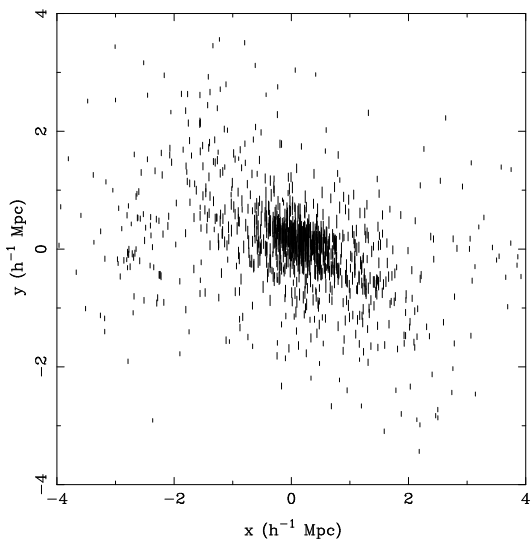}
}
\caption{ The  effect of gravitational  redshift for the  most massive
cluster. The line  of sight lies parallel to the y-axis.
 The  lines start at the
position of a galaxy before  adding its gravitational redshift and end
at  the  position   including  the  gravitational  redshift.  Peculiar
velocities are not added in the plot.
\label{fig:cgz}
}
\end{figure}

In  Figure~\ref{fig:cgz} we show the effect of 
gravitational redshifts on the positions of particles
when they are plotted in redshift space. 
We choose $y$ as the line of sight.In this figure
 we have not included the
effect of peculiar velocities so that the much smaller
gravitational redshift effect can be clearly seen.
The value of $\Delta v_{g}  $ for the central galaxy is $\sim 40
km/s$, which causes it to 
be displaced $0.4 \hmpc$ further
away from the observer than in real space.
The gravitational redshift gets  smaller rapidly as we move
away from the center of the cluster and particles at the edges of the
cluster mass distribution have $\Delta v_{g}$ values wrt infinity
of $\sim 5 \kms$.

\section{Results}
\subsection{Comparison with an Analytic Profile}
We  compare  the  gravitational  redshift  from  the  simulation  with
an analytic prediction using a spherically symmetric density profile.
 We 
are interested in the difference in
gravitational  redshift  between  the central galaxy  and  the  other
cluster  members and we will compare our analytic results with those from
the simulation. 
The  density profile we adopt is
given by (Hernquist 1990):
\begin{equation}
\rho (r) \propto \frac{1}{r(a+r)^{3}}
\label{eqn:hern_d}
\end{equation}
where $a \sim R_{e}/1.8153$.
The spatial gravitational redshift profile implied by this is (Cappi 1995):
\begin{equation}
V_{H} = \frac{G M}{c(r+a)}.
\label{eqn:hern_grav}
\end{equation}
The gravitational redshift of the central galaxy 
or of the cluster is then given by integrating the 
mass weighted $V_{H}$ value over the radius of the galaxy or the
radius of the cluster, as in the following expression:
\begin{equation}
v_{g} = \frac{\int^{r_{g}}_{0} \rho (r) V_{H} (r) dV}{\int^{r_{g}}_{0}
 \rho (r) dV} 
      = \frac {G M}{c} \frac{\int^{r_g}_{0} \frac{r}{(a+r)^{4}} dr}
{\int^{r_g}_{0} \frac{r}{(a+r)^{3}} dr},
\label{eqn:anal_grav}
\end{equation}
where here $r_{g}$ is the galaxy radius.

We  use  the  trapezoidal  rule to  integrate  Eq.~\ref{eqn:anal_grav}
numerically. For  the 
characteristic  radius  of  the
central galaxies  and the  clusters, we  take the
mass weighted mean
radius  of galaxy (or  cluster) member particles.  With this  definition, the
average characteristic radius of  clusters in the simulation 
with mass greater than  $10^{13} \hmpc$ is  0.338 $\hmpc$,
and the average radius of galaxies in those clusters is $0.084 \hmpc$
With a threshold in mass of $10^{14} \hmpc$ the average
radius of clusters and galaxies respectively is $0.713 \hmpc$
and $0.154 \hmpc$. The
characteristic radius of the largest cluster in simulation is 1.9 $\hmpc$.

\begin{figure}[t]
\centering
\vspace{0.cm}\psfig{file=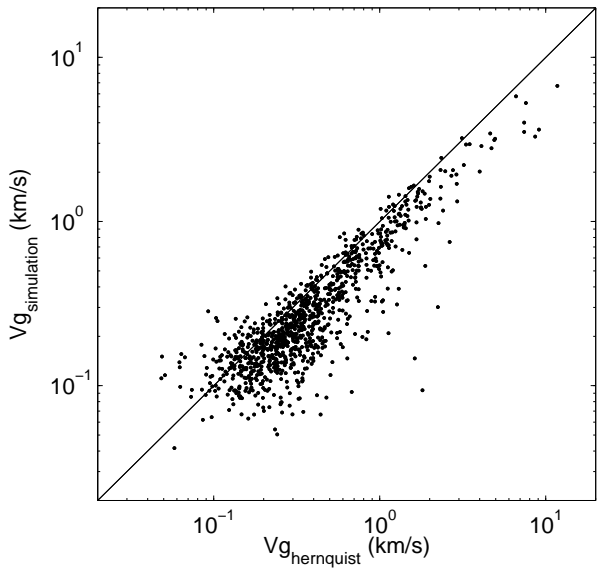,angle=0.,width=8.truecm}
\caption{  Difference  in  gravitational  redshift  between  the  most
massive  galaxy  and the  rest  of  cluster  members. $x$  axis:  from
Hernquist  analytic profile  using  Eq~\ref{eqn:anal_grav}, $y$  axis:
from  simulation.  The  straight   line  is  where  $Vg_{hernquist}  =
Vg_{simulation}$.
\label{fig:hern_sim}
}
\end{figure}
For   the  gravitational   redshift  of   the  most   massive  galaxy,
Eq.~\ref{eqn:anal_grav} has been integrated up to the radius of the most
massive  galaxy, and  for the  gravitational redshift  of the  rest of
members,  the size  of the  most massive  galaxy and  the radius  of the
cluster  are  used  as  the  lower and  upper  limits  of  integration
respectively. The difference between the two values are calculated and
the results  from the simulation and  from Eq.~\ref{eqn:anal_grav} are
plotted in  Figure~\ref{fig:hern_sim}.
We can see from the figure   that the values of 
central galaxy gravitational  redshift from  the  simulation 
are broadly  consistent with  the
estimate from analytic Hernquist density profile,
although the estimate from analytic profile is slightly larger (by 
about $25 \%$) 
than the result from the numerical simulation. We investigated if 
this could be due to the substructure of clusters, i.e., we checked
whether $\Delta v_{g}$ 
is large when $\Delta y_{c.m.}$ is small, which might be 
expected if concentricity of the galaxy and cluster yields a
greater gravitational redshift. We find no such effect, however.
A possible explanation of the difference could follow from the 
known fact that the density profiles of halos are better fitted
by the NFW
profile which gives $\rho \propto 1/r^{3}$ for large $r$ rather than
the steeper ($\rho \propto 1/r^{4}$) Hernquist 
profile. The latter would lead to a greater difference in the potential of the
outer and central parts. 
In any case, the scatter  we see  in 
Figure \ref{fig:hern_sim} is quite small, indicating that substructure
and deviations from spherical symmetry do not 
disrupt the agreement with the analytic results. 
Also apparent from the  
plot is the range of gravitational redshifts which will be relevant in
our study, from $\sim 1 \kms$ to $\sim 10 \kms$.

\begin{figure}[b]
\centering
\psfig{file=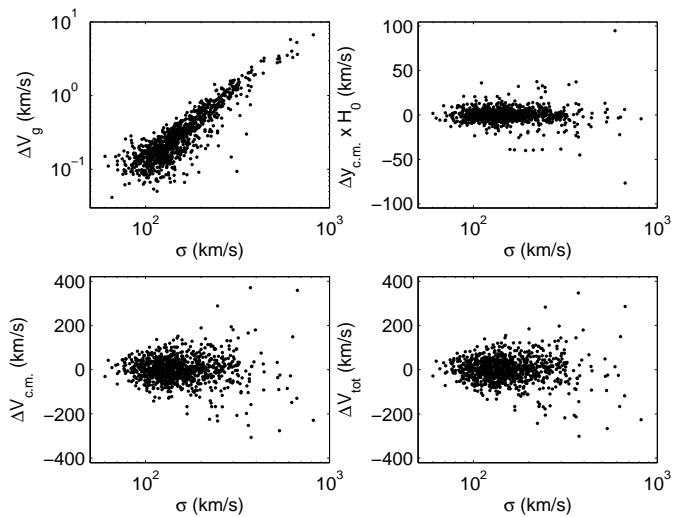,angle=0.,width=9.truecm}
\caption{  The  difference  in  gravitational redshift  velocity  (top
left), center  of mass  (top right),  and center  of mass  velocity (bottom
left) of the central galaxy from  the rest of the cluster members as a
function  of velocity  dispersion. The  bottom right  panel  shows the
total redshift velocity, which is the sum of the
3 previous components. 1,000 randomly chosen clusters are shown.
\label{fig:dvtot}
}
\end{figure}

\subsection{Gravitational Redshift in Redshift Space}
The total redshift velocity difference we observe between a galaxy and 
cluster 
 is given by:
\begin{equation}
\Delta v_{tot} = \Delta v_{g} + \Delta y \times H_{0} + \Delta v ,
\label{eqn:vtot}
\end{equation}
where $\Delta y  \times H_{0}$ is the Hubble  expansion part and $\Delta v$
is the  peculiar velocity. For each  component,
 we have calculated the difference in redshift
between the most massive galaxy and  the other cluster
members (see also \S3.3). There will also be
measurement errors on the redshift, 
which we shall cover later. In Figure~\ref{fig:dvtot} we plot each redshift
component separately and  the  total  redshift  as  a 
function  of  the cluster line-of-sight  velocity
dispersion. In order that these
values may be related to cluster mass, we 
have fitted a power law to the relationship  between  cluster  mass and
velocity dispersion, finding it to be:
\begin{equation}
\log M = (2.56 \pm 0.04) \log \sigma + 7.12 \pm 0.08,
\label{eqn:m_sig}
\end{equation}
where $M$ is in units of $\msun$ and the units of $\sigma$ are $\kms$.
The slope of the relation is
not too far from the value of 2 expected for virialized objects.
The massive, high velocity dispersion clusters  have the largest
central galaxy gravitational redshifts.  The rest  of the redshift
components exhibit larger scatter as velocity dispersion
increases.

The cluster  in the  bottom right panel  in Figure~\ref{fig:clusterxy}
has more substructure  than the others and this  is reflected in the large
redshift difference between the  most massive galaxy and the  center of mass of
rest of members.
The position differences for all the panels  are 
$\Delta  y_{c.m.}   =$  0.0698,   0.049,  -0.181  and   1.309  $\hmpc$
respectively    (clockwise    from    the    top   left      in
Figure~\ref{fig:clusterxy}).  Compared  to  the  velocity  difference,
however,     this     has    a small     effect     on     the     total
redshift
(see top right  panel in Figure ~\ref{fig:dvtot}).
  Figure~\ref{fig:dvtot} shows  that   most  scatter comes
from the velocity difference between the central galaxy and the 
rest of the cluster.

The plot  shows how small  the gravitational redshift is  compared to
the  other redshift components. In the bottom right panel 
of Figure \ref{fig:dvtot} where we show the
total redshift, it is obvious that we cannot see the effect of
gravitational redshifts by eye on individual clusters
or overall in a scatter plot such as this one.

\begin{figure}
\centering
\psfig{file=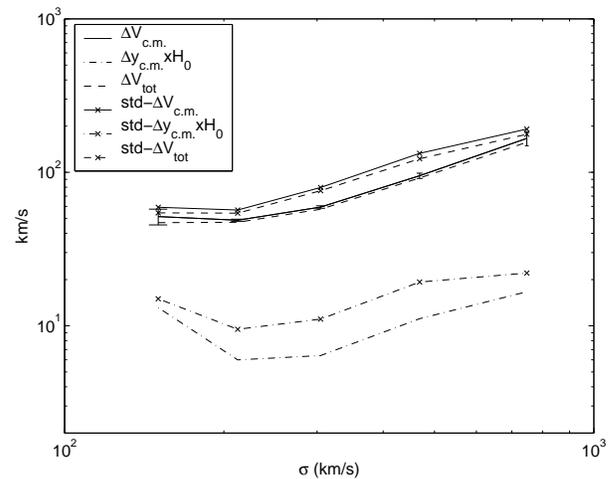,angle=0.,width=8.truecm}
\caption{ Difference in redshift between central galaxy and cluster 
centroid as a function of cluster velocity
dispersion $\sigma$. We show the scatter in
center of mass velocity, Hubble component, and total
redshift, as in Figure 5.
The lines show the values which  enclose 68\% of the distribution
 and symbols the $1\sigma$ 
standard deviations.   Error bars (Poisson errors) are shown for the center
of mass velocity.
\label{fig:sig68}
}
\end{figure}

  The redshift due to the position  and
velocity differences will be the most important source of
noise when  we try to extract  the gravitational redshift. 
If the most massive galaxy were lying at rest at the centroid of the 
cluster, then the accuracy of our measurement of  the gravitational redshift 
difference between it and the other cluster members would be limited
only by Poisson noise in computation of the mean redshift of the cluster.
Because the central galaxy is usually moving with respect to the
mean of the other cluster members, and is usually
not at the cluster  center of mass, we have this additional
source of noise, shown in the bottom left panel of
Figure 5. Averaging results over many clusters in order to
reduce the error on the mean  \vg is the approach we will take 
in this paper. An interesting question is how best to do this.
 For  the 
clusters with the greatest mass, \vg  will be  large, but so will the
 noise per cluster. On  the other
hand, \vg for  small mass clusters is small but the peculiar
velocity and position noise is small as
well. We will discuss this more later in  the paper and 
determine the  optimal range of
cluster mass for detection of gravitational redshifts.

One can ask how one might expect the dominant noise component 
$\Delta V_{c.m.}$ to scale with the cluster velocity dispersion $\sigma$.
If the central galaxy were a random galaxy in the cluster, then the 
scatter in $\Delta V_{c.m.}$ should be approximately $\sigma$. However,
 we expect it to be substantially less than this, as the central galaxy,
while not at rest is still moving less with respect to the cluster center
of mass than a random galaxy. In order to quantify this, we have plotted in
Figure  \ref{fig:sig68} the standard deviation of $\Delta V_{c.m.}$
against $\sigma$. We find standard deviation of $\Delta V_{c.m.}$
to be approximately $0.35 \sigma$. We have also plotted the other source
of noise from the Hubble component
$\Delta y_{c.m.}\times H_{0}$, as well as the total,
$\Delta V_{tot}$. Interestingly,
 the total noise on the central galaxy position,
$\Delta V_{tot}$ is slightly lower than
$\Delta V_{c.m.}$. On investigation, we have found that this is because of
residual coherent infall of particles into the cluster, so that 
clusters on the far side, with +ve $\Delta y_{c.m.}\times H_{0}$ have
-ve values of $\Delta V_{c.m.}$ and vice versa. In Figure 6,
 we also compare the directly computed
 $68 \%$ interval to the standard deviation, finding that
the distribution of $\Delta V_{tot}$ is close to Gaussian.

\begin{figure}
\centerline{
\psfig{file=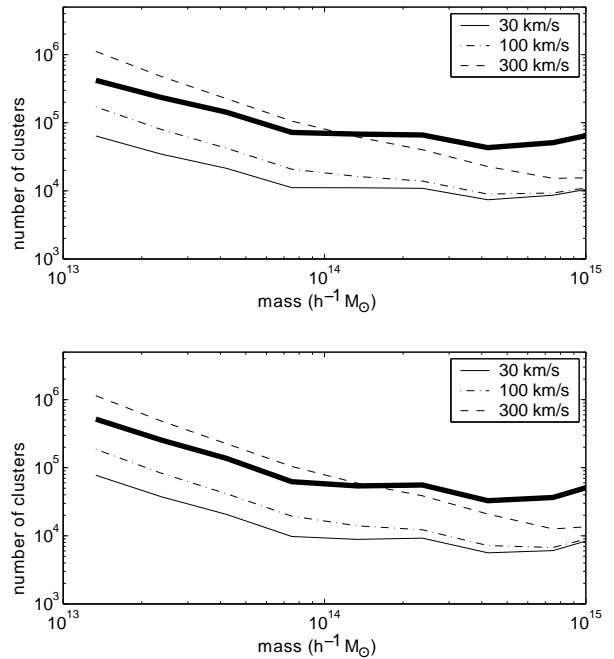,angle=0.,width=8.truecm}
}
\caption{ The number  of clusters  needed
to detect graviational redshift as a  function of cluster mass.
We show results using either 50
randomly selected  galaxies per cluster (top) or using a number
of galaxies proportional to the cluster mass (bottom), both for 
a 2$\sigma$ level detection.  For the latter case,
 the number of galaxies used is equal to 0.1 times the number of
dark matter particles in it, so that for a $10^{13} \msun$ we use
15 galaxies. Measurement uncertainties 
of 30 km/s, 100 km/s and 300 km/s are included. The thick line represents
 a 4$\sigma$
detection with no measurement uncertainty. Note that the graph flattens
for large cluster masses.
\label{fig:nc_2bin}
}
\end{figure}

\subsection{Estimate of the Number of Clusters Needed to Observe
 Gravitational  Redshifts}

One simple
method for detection of gravitational redshifts would involve finding the
redshift difference between the central galaxy and the other cluster
members, after averaging over a large number of clusters. We
have seen in the previous section that the noise from substructure
is small, and that this should be feasible given enough clusters.
In this subsection we investigate how well this could be done in practice.
In the next subsection, we will include information on all galaxies
in the cluster rather than just splitting the galaxy population
into ``central galaxy'' and ``other''.

  With  the  difference   in  gravitational
redshift computed from the simulation,
 we make an estimate of the number of clusters needed
for a detection. We try two cases, randomly selecting
either $N_{gal}=50$ galaxies per cluster 
or $N_{gal}$ = one tenth of the number of dark matter particles,
so that $N_{gal}=M_{cluster}/7\times 10^{11} \msun$.
We assume
that  the  noise  from  the  difference  in  position  and  velocities
contributes independently. 
Including the measurement uncertainty $\Delta_{meas.}$, we have
\begin{equation}
\sigma_{tot}^2 = \sigma_{vel.}^{2} + \sigma_{pos.}^{2} + \Delta_{meas.}^{2},
\label{eqn:sig2}
\end{equation}
where 
\begin{equation}
\sigma_{vel}^{2} = \frac{1}{N_{c}-1}\sum_{i=1}^{N_{c}} (v_{i} - v_{gal,i})^2
\label{eqn:sig_v2}
\end{equation}
and 
\begin{equation}
\sigma_{pos.}^{2} = \frac{1}{N_{c}-1}\sum_{i=1}^{N_{c}} (r_{i} - r_{gal,i})^2
 \times H_{0}^2
\label{eqn:sig_p2}
\end{equation}
And the error on the mean is given by 
\begin{equation}
\frac{\sigma_{tot}}{\sqrt{N_{0}}}, 
\label{eqn:error}
\end{equation}

We have split the clusters in the
simulation into bins by mass,
and  $N_{c}$ here is  the number of clusters in the simulation
in a particular bin.
For each bin
we compute the number of clusters ($N_{0}$) actually needed for
a detection
of the gravitational redshift, with a given statistical significance.
  For  example, for clusters in
the mass bin centered on $\sim  8 \times
10^{14}  \msun$,  $\sigma  \sim  300  \kms$ and  $\Delta  v_{g}  \sim7 
\kms$. With $\sim10^4$ clusters 
we expect our error bar to be $\sim3 \kms$,
which is a $\sim 2 \sigma$ detection of gravitational redshift.

We might expect that  the number of  clusters should decrease for
high mass  because the gravitational  redshift is proportional  to the
cluster velocity dispersion.
  However, for  these clusters,  the error  bars also
increase    because    the   dispersion in  velocity  
difference between central galaxy and others   is    getting
larger. Figure~\ref{fig:nc_2bin}  shows how  many clusters we  need to
detect the  gravitational redshift at the 2$\sigma$  and 4$\sigma$ levels
as a function of cluster mass. We choose the measurement uncertainty 
to be 30 km/s, 100 km/s and 300 km/s
 as best, typical and worst cases (e.g., Stoughton et al. 2002). 
Given relatively small measurement uncertainty, the number of clusters
needed decreases
at first  and then  stays roughly  the same  ($\sim10^4$  clusters for $\Delta_{meas.} =$ 30 km/s) after $M \simgt 10^{14}\msun$. Since we use average 
values, it does not make much difference whether we use 50 galaxies per
cluster or a number proportional to the cluster mass. 

In order to understand the behaviour seen Fig \ref{fig:nc_2bin}, we
can ask what should happen if
 the redshift noise on the central galaxy from 
peculiar velocities and position differences were proportional 
to the cluster velocity dispersion (assuming 
we are averaging over a fixed number of galaxies in each cluster).
As the gravitational redshift is proportional to cluster 
velocity dispersion also (Figure \ref{fig:dvtot}),
 then the number of clusters for a detection
of a given significance should be constant. This is 
approximately what is seen for clusters of mass $M$ greater than 
$\sim 10^{14}\msun$. For smaller clusters, 
the noise from the displacement of the  central galaxy (the Hubble
velocity component, $\Delta y_{c.m.}\times H_{0}$) is becoming
a non negligible fraction of the total noise, and its effect means that 
more clusters must be averaged over.

 Note  that, as mentioned above, only the mass-averaged gravitational
redshift has  been used here  and it has  not been taken  into account
that the  gravitational redshift has a gradual radial dependence. More 
information is therefore available which could in principle aid a detection.
We discuss this in the following subsection.

\subsection{Gravitational Redshift Profile} 

We now calculate the profile of gravitational redshifts
averaged over many clusters. As in the previous section, we 
take the central galaxy to be our zero point. The other galaxies 
in the cluster are binned as a function of their
impact parameter from this galaxy ($r = \sqrt{\Delta x^{2} +
\Delta z^{2}}$, where $y$ is the line of sight)
In each bin, and for all clusters, we have summed the gravitational redshift
difference between the galaxy and the central cluster
galaxy. We then divide by the number of galaxies
in each bin (as mentioned previously the ``galaxies'' which are
not the central one are actually particles). 

\begin{figure}[b]
\centering
\vspace{0.cm}\psfig{file=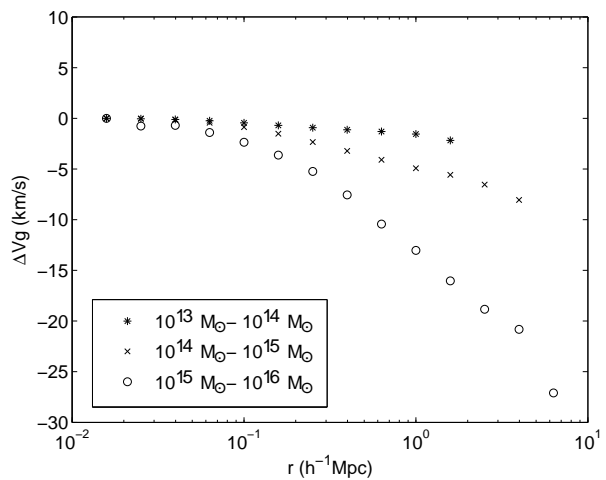,angle=0.,width=8.truecm}
\caption{  The  gravitational  redshift  profile  averaged  over  many
clusters  as a  function of  impact  parameter from  the most  massive
galaxy. The gravitational redshift of the central galaxy is subtracted
from  the  average  gravitational  redshift  of the  rest  of  cluster
members. 
\label{fig:dvg}
}
\end{figure}

Figure~\ref{fig:dvg} shows  the resulting gravitational
redshift profile for clusters within different
mass ranges as a function of the impact
parameter $r$. All of the clusters in  the simulation  have been
used, and  it should be  noted that each  mass bin does not an have equal
number of clusters contributing to it. 
Also, at large radii, within each mass range,
only the more massive clusters contribute, because the smaller
clusters do not have any galaxies out that far. As  expected,
gravitational redshifts increase for  large mass clusters. This result
is  consistent with that of Cappi (1995) who used analytic
profiles,  in that  the function  decreases smoothly
with a maximum at the center.

As we have described in section 3.1 our central galaxies have a mean radius of
$\sim 0.1 \hmpc$, which means that within the central few bins, we do not 
expect the gravitational redshifts of the galaxy to rise very steeply.
If the central galaxy were modelled as a point (as was done by
Broadhurst \& Scannapieco 2000), the profile could become steeper.  The 
dark matter halos in this simulation will have profiles 
consistent with the NFW form, so that we do not
expect the central profile to be very steep even if the central
galaxy is allowed to be much smaller, at least in this
dark matter only run.  The gravitational 
redshift profile for NFW clusters is shown in Broadhurst \& Scannapieco 2000.,
and flattens off towards the center. 
According to El-Zant et al. (2003), any baryonic dissipational effects 
can cause either a steeping or
flattening of the density profile towards the center, which will 
increase or decrease the gravitational redshift, but this is 
ignored in the present study.

In order to use the redshift profile information to make a
detection of gravitational redshift, we must average over a
large number of clusters to reduce the noise from  peculiar
velocities and Hubble expansion.
As before,  we  add the total redshifts for  many clusters so
that  the  redshifts from  position  and  velocity  are averaged  out,
leaving only the gravitational redshift. This time, this will be
done in bins of impact parameter.
We can then estimate how many clusters we need to observe in
order to ascertain to a given significance level that the gravitational
redshift is inconsistent with zero.

\begin{figure}[t]
\centering
\vspace{0.cm}\psfig{file=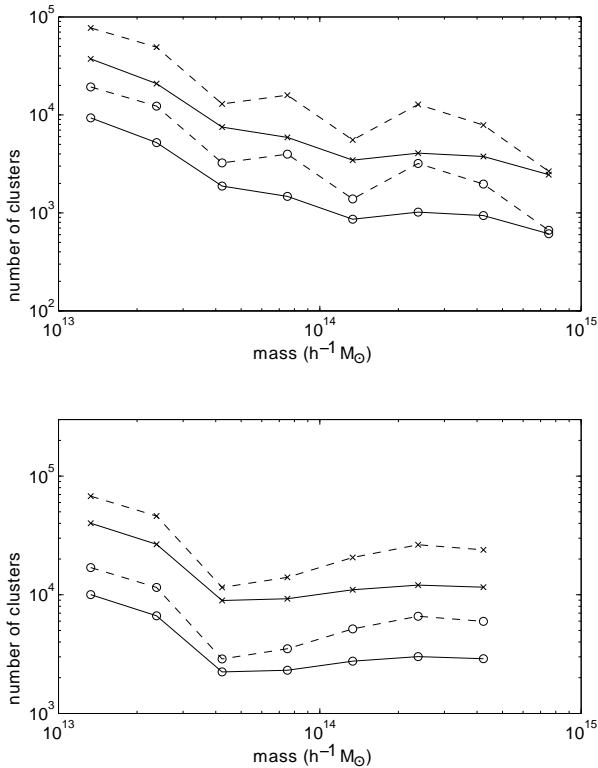,angle=0.,width=8.truecm}
\caption{ top: Number  of clusters needed to  detect gravitational redshifts
at  the  $2\sigma$  (circle)  and  $4\sigma$ ($\times$)  levels  as  a
function  of cluster mass.  
The $\chi^{2}$  has been  calculated  with (dotted
line) and without the (solid line) covariance matrix. We use all particles
in each  cluster. bottom: Number of clusters needed 
to detect gravitational redshifts
as  a function of mass when we are
restricted to  100 galaxies per cluster. 
Note that  the graph  flattens for  the mass
above  $\sim 3  \times 10^{13}  h^{-1} M_{\odot}$  when  only diagonal
terms are used. Measurement uncertainties are not included.
\label{fig:nc_all}
}
\end{figure}

First,  given a sample of simulated clusters in
a certain mass range, we calculate the gravitational redshift profile
in bins of impact parameter and the matrix of covariance between
bins. 
The technique of Singular Value Decomposition (Press et al. 1992) is
used to invert the matrix, and then the value of $\chi^{2}$ is
computed:
\begin{equation}
\chi^{2} = \sum_{N} (x_{i} - \overline x_{i})  
C_{ij}^{-1}  (x_{j} - \overline x_{j})
\label{eqn:chi2}
\end{equation}
where  $x_{i}$ is the gravitational  redshift, $C_{ij}$ is covariance matrix.
As we are interested in the possibility of detection, we 
set $\overline{x}_{i} = \overline{x}_{j}=0$ exactly so that $\chi^{2}$ 
is then the difference $\Delta\chi^{2}$ from a model with zero
gravitational redshift. Because $C_{ij}$ is proportional
to $1/\sqrt{N_{i}N_{j}}$, where $N_{i}$ is the number of
clusters contributing to $r$ bin $i$,  $\chi^{2}$ is
proportional to the total number of clusters, if $N_{i}=N_{j}$.
 The number of $r$ bins
which could be used depends on the angular extent of the cluster.
 Our criterion is  to truncate at the outermost bin with less than 20
clusters contributing and not  use the larger $r$ bins to 
 build the  covariance
matrix.  Additionally, when the number of clusters in a given mass
range is small, the matrix is too noisy to invert.
We  consider  the results invalid  when  $\chi^{2}$
calculated with the covariance matrix is greater than $\chi^{2}$ computed
using only the  diagonal elements of the covariance matrix. 
This occurs for the clusters with  $M >  5
\times   10^{14}   \msun$   since    there   are   only   4   in   the
simulation, and for this mass range we use the diagonal elements
of the matrix only.

The top panel in Figure~\ref{fig:nc_all}  shows  the  number  of  clusters
needed for detections of non-zero gravitational redshifts at the
2$\sigma$  and 4$\sigma$ 
levels as a function of mass. No measurement 
uncertainties are included in this plot. $\chi
^{2}$ values with only  the diagonal elements were also  calculated and 
those results are shown
as additional lines in the plot. Including the proper covariances
between bins reduces the sensitivity of the test, so that
approximately a factor 2 more clusters are needed.  The plot has 
slightly different behaviour to Figure \ref{fig:nc_2bin}, as the 
number of clusters needed continues to decrease for high masses.
This is basically because we have used all galaxy information
in each cluster. This is obviously not very realistic, as 
up to 20000 galaxies (particles) have been averaged over in the most 
massive clusters. We have therefore recomputed our results assuming
that only 100 galaxies per cluster are available.

When we do this as shown in the bottom panel in Figure~\ref{fig:nc_all},
 we find that 
the curve flattens for high mass clusters and even goes up slightly
when the effect of covariances is included. 
This means  that observing clusters of mass
$\sim 5 \times 10^{13} \msun$ seems to be the most efficient way to 
detect gravitational
redshifts, at least when using galaxies as tracers of the potential.
  For higher  mass clusters,  we have  larger gravitational
redshift effects but also  have more noise. 

Note that when only the diagonal terms in the covariance
matrix are used  in $\chi ^{2}$ calculation  the flattening of
the graph is    consistent    with    the  simpler    estimator     in    \S3.2
(Figure~\ref{fig:nc_2bin}). However,  both the mass threshold and the number
of  clusters  needed to detect gravitational redshifts
  are  smaller here, by about
a factor of 3. This is  because we  have  more
information in the sense that we use the gravitational redshift profile
in many  $r$ bins  while in \S3.2,  there are  effectively only two bins  - the
central galaxy  and the  rest of the galaxies in each cluster.

\begin{figure}[h]
\centering
\vspace{0.cm}\psfig{file=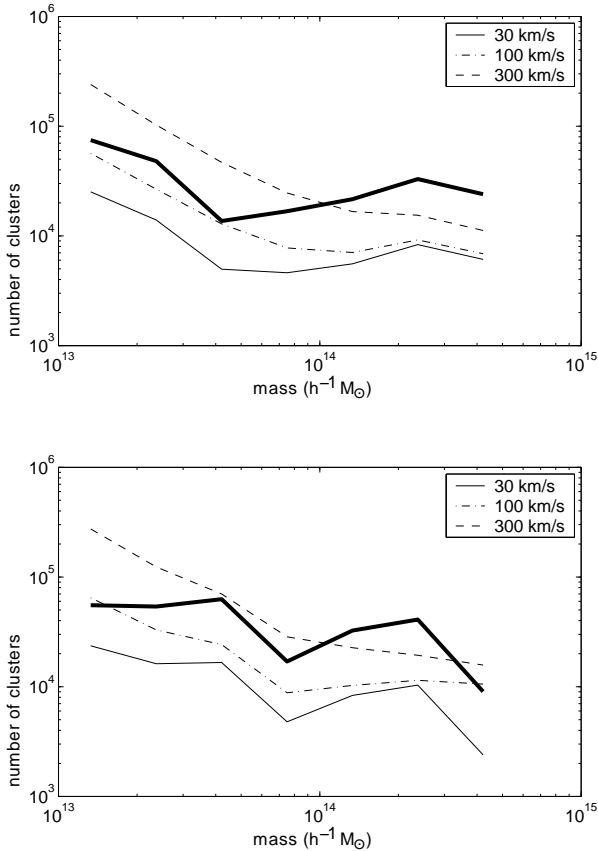,angle=0.,width=8.truecm}
\caption{ top: Number  of clusters needed to  detect gravitational redshifts
at  the  $2\sigma$  level  as  a function  of cluster mass.  
The measurement uncertainties of 30 km/s, 100 km/s and 300 km/s are included.  
We use 50  particles per cluster (top) and tenth the size of cluster (bottom).
Note that  the graph  flattens for  the mass
above  $\sim 3  \times 10^{13}  h^{-1} M_{\odot}$. The thick line 
represents the  $4\sigma$  level with no measurement uncertainty.
\label{fig:nc_50_tenth2}
}
\end{figure}

For a more direct 
comparison with Figure 7, 
we then use 50 galaxies per cluster and one-tenth the number of
particles while including  measurement uncertainties in the covariance matrix 
(Figure~\ref{fig:nc_50_tenth2}). When the uncertainty is small, the 
result is consistent with Figure~\ref{fig:nc_2bin}
in that with fixed number of galaxies per cluster, the curve flattens
for high mass clusters. However, when the uncertainty is large, 
it appears that the measurement
 uncertainty is the dominant factor deciding the number 
of clusters needed  
regardless of the number of galaxies per cluster. 

We have seen that for a $2 \sigma$ detection of gravitational redshifts
we expect to need a sample of $\sim 5000$ galaxy clusters/groups with
50 measured redshifts per cluster
 ( with measurement error $\Delta_{meas.}$ = 30 km/s).
A statistically correct detection
would involve making use of the covariance matrix of the noise between
bins, which could be done from our model clusters, or else computed 
directly from the observational sample itself. This 
required number of clusters/groups is large, but may be 
just within the range of
present day redshift surveys. We will return to this in our discussion (\S 5),
as well as exploring results from a recent survey below.
Alternatively, as stressed by Cappi (1995), the observed absence of an effect
could be used to place limits on the gravitating mass of clusters.

\section{Results from the ENACS survey}
The  ESO Nearby  Abell  Cluster  Survey (ENACS,  Katgert  \etal 1998) is  a
redshift  survey of  rich  Abell  clusters in  the  southern sky.  The
catalog  contains redshift,  magnitude and  position information
for 5634  galaxies in  107 clusters. The  mean velocity  dispersion of
these clusters is $\sim 700 \kms$  (Biviano et al 1996), so that based
on the  work in \S  3 we only  expect there to  be a small  ($\sim 5
\kms$) redshift  difference on average 
between the  central galaxy and  the other
cluster  members. We  will compute  this statistic  in order  to check
consistency with our expectations.

\begin{figure}
\centerline{
\psfig{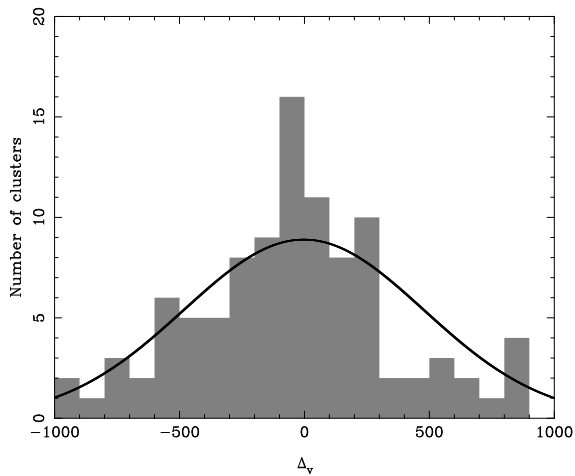}
}
\caption{ 
Difference in redshift between the brightest cluster galaxy and the
other cluster members, for 107 rich Abell clusters from the
ENACS. A Gaussian curve with the same standard
deviation, $\sigma=486 \kms$ is also shown. 
\label{clusterz}
}
\end{figure}

We take the ENACS data and find cluster centers
by placing a cylinder of length $3000 \kms$ along the line of sight
and radius $2.0 \hmpc$ around each galaxy. We find the 
center of mass of each cylinder (weighting
each galaxy by the inverse of the selection function),
 recenter it on this point
and iterate until convergence. We eliminate all overlapping cylinders,
keeping those with most galaxies. We then rank all cylinders by the 
weighted number of galaxies and end up with a list
of 107 rich clusters and their members.

 Because we do not expect to be able to detect a gravitational redshift signal,
we only carry out the simplest test, measuring the
difference in redshift between the brightest galaxy in each cluster
and the other cluster members. The results are shown as a
histogram in Figure \ref{clusterz}. The $rms$ difference between the brightest
and other cluster galaxies is $486 \pm 47\kms$ (Poisson error). 
In the simulations, for a sample
of clusters with the same mean velocity dispersion we
expect an rms difference of $255 \kms$ (from Hubble flow and 
peculiar velocities.)
This is rather lower than the observed value, and in fact it takes
a simulated sample with a  mean
velocity dispersion of $970 \kms$ to have the same $rms$ difference
(by applying a lower mass threshold of $9.3 \times 10^{14} \msun$. This
could be a sign that there may some differences between the structure of
observed and simulated clusters. It is 
likely  that this is at least partly due to observational measurement 
errors, as the quoted uncertainties of $\sim 100 \kms$, when 
added in quadrature would increase the rms difference to $275 \kms$. 
 The mean of the histogram for the ENACS clusters is $-66 \kms$ (a blue shift),
with a $1\sigma$ uncertainty of $47 \kms$ so that
as expected there is no detectable evidence for gravitational redshifts from
this small sample.

We note that Cappi (1995) also carried out a similar
 same test with a sample of 
42 clusters with cD galaxies, and found a small blue shift, 
$-39 \pm 32$ km/s. Although in the ENACS sample we have simply measured
the redshift difference of the brightest galaxy without regard to
whether it is a true cD or not, we find essentially consistent results.
On the other hand, Broadhurst and Scannapieco (2000) find a mean cD
redshift of $260 \pm 58 \kms$ for a sample of 8 clusters, of which 5 have
cooling flows. This value is an order of magnitude larger than
what we expect from our simulations. These authors have pointed out
that the central potential could have been made steeper over time by
cooling flows.

\section{Summary and Discussion}
We have explored the gravitational redshifts of galaxies in the 
potential wells of simulated galaxy clusters. In particular
we have concentrated on the possibility of detecting gravitational
redshifts from surveys of cluster galaxies. Our main findings
are as follows:

\begin{itemize}
 \item For clusters of  $\sim10^{15}\msun$, the difference in redshift
 between the central galaxy  and the  other cluster  members is  $\sim 10
 \kms$.
\item The gravitational potential of clusters does not exhibit much
 substructure, even when the clusters are not relaxed. The most 
important source of noise on a measurement is from galaxy 
peculiar velocities.  
\item The  most sensitive method for detecting gravitational
redshifts with galaxy data  is to use the
 gravitational redshift profile as a function of 
impact parameter.  
 \item For a given number of galaxy redshifts measured, the most
 efficient  strategy would  be  to target  clusters  of $\sim5  \times
 10^{13}\msun$, which have lower noise from galaxy peculiar
velocities than more massive clusters.
 \item For a 2$\sigma$ detection, we require $\sim 5000$
 clusters with  $M >  5 \times 10^{13} \msun$ when the measurement 
uncertainty is 30 km/s and $\sim 2500$ clusters when the measurement
uncertainty is negligible.
\end{itemize}

Our result for the shape of the gravitational  redshift profile are 
 consistent with
that of Cappi(1995) and Broadhurst and Scannapieco (200).
 However, the masses of clusters in our simulation are generally smaller than
those used as examples by these authors.
For example Cappi (1995) found that gravitational redshift with
respect to infinity of the central regions of a
cluster of mass $M  = 5 \times 10^{15} \msun$  to be $> 100 \kms$
and  predicted the  effect  for even  more massive clusters  with $M  =
10^{16}    \msun$ ($v_{g}  \sim  300  \kms$). Unfortunately, 
even though the gravitational redshift in a cluster
with mass $10^{16} \msun$ is large,
there is little chance of finding such a massive cluster  in the
 Universe. 

We can  roughly estimate the number of clusters  in the universe whose
mass is $M = 8 \times 10^{14}  \msun$ or greater.
First, we approximate the mass function of clusters
given by (Bahcall \& Bode, 2003) as a power law:
\begin{equation}
\log n (>M_{1.5,com}=8\times10^{14} \msun) \approx -2 z - 6.75,
\label{eqn:bahcall}
\end{equation}
The comoving distance as a function of redshift is given by (e.g., Hogg 1999):
\begin{equation}
D = D_{H} \int^{z}_{0} \frac{dz'}{E(z')}
\label{eqn:hogg}
\end{equation}
where $E(z) \equiv \sqrt{\Omega_{M} (1+z)^{3} + \Omega_{k} (1+z)^{2} +
\Omega_{\Lambda}}$  and  $D_{H} =  3000  \hmpc$. We take
$\Omega_{M} = 0.3$ and $\Omega_{\Lambda} = 0.7.$ We then find
 the total number
of  clusters in the universe (up to $z=1$, although
raising the $z$ limit makes essentially no difference) :
\begin{equation}
N = \int n dV = 4 \pi D_{H}^{3} \times 10^{-6.75} \int^{1}_{0} 
\frac{10^{-2 z}}{E(z)} \left( \int^{z}_{0} \frac{dz'}{E(z')} \right) ^{2} dz.
\label{eqn:totn}
\end{equation}
Integrating Eq.~\ref{eqn:totn} numerically  we find that there are
only $\sim 600$ clusters ($M >  8 \times 10^{14} \msun$) 
observable in the
entire universe. We have seen, however in \S3, that in 
any case the larger noise from peculiar velocities in 
more massive clusters makes a detection of gravitational redshifts
 problematic anyway.
It might  be  better to  use lower  mass
clusters in  order to detect  the gravitational redshift, even though
effect is small.

As for observed samples of clusters with lower masses, the situation
is  more promising. For example, the Sloan Digital Sky
Survey has been used to find $\sim  1,000$ clusters in 
 total with $z < 0.15$ and  $M> 5  \times 10^{13} \msun$ so
far (Chris Miller, 2003, private communication). For 
a more on cluster finding in the SDSS, see Nichol (2003) and Annis \etal
(2003).
Published samples of 
SDSS clusters include those of Goto \etal (2002) and Bahcall \etal (2003b). 
 Merchan and Zandivarez (2002)
have constructed a galaxy group catalog from the 100K data release of the
the 2dF  galaxy  redshift survey (refs). They find  $\sim  1000$
clusters/groups  with masses  $M> 5  \times 10^{13}  \msun$.  
The final 2dF data release
contains $\sim  2 \times  10^{5}$ galaxies  (Colless et
al. 2003), which should enable construction of a group/cluster
catalog with $\sim 2500$ members above this lower mass limit.
As we have seen in \S 3.4, this may be close to the level required in order to 
make a $2 \sigma$ detection of the gravitational redshift. A catalog of 
similar groups selected from the final SDSS survey might allow a 4 $\sigma$
detection.

Because the effect we are searching for is very small (a
redshift on the order of a few km/s), we must be careful to consider
possible systematic effects.  The effects of peculiar velocities and
cluster substructure have been included in our simulations, and
averaging over large numbers of clusters produces reasonable results.
One effect we have not included is full selection
of clusters in redshift space. In the simulation, we  simply define a 
 cluster in real space with  a linking
length of our choice. In  a survey, we select the clusters in redshift
space and we  include  some contaminating galaxies  behind and 
in front of each cluster that are not members.
 This would act to increase the noise, but not by much, as they
 will be heavily 
outnumbered by cluster members.

A real problem would be any effect which gives a systematically
positive or negative redshift for the central galaxy with respect to
the other galaxies. If we select cluster galaxies using a 
cone of constant angular size rather than a cylinder of fixed radius,
then more contaminating galaxies will come from the larger volume 
behind the cluster than in front, and the average redshift
of the galaxies apart from the central one will change. We
have calculated the potential
effect of this bias using the cluster-galaxy cross-correlation
function data of Croft \etal (1999) to find the number
of contaminating galaxies. We find a relative blueshift of $\sim 1\kms$
for the central galaxy. This can be avoided simply by using a cylinder
to select cluster members, however.

An additional potential problem which goes in the
opposite direction is the fact that with a magnitude limited
survey, the contaminating galaxies behind the cluster should be fewer
in number because they are at a greater distance. The size of this
effect depends on the steepness of the
luminosity function of galaxies. It could also however easily be 
eliminated by applying a local volume limit in the vicinity of each 
cluster, rejecting galaxies which would be too faint when placed at the
far edge of the cluster. 

In this paper we have seen that realistically, gravitational redshifts
are an extremely small effect, and that looking at the most massive
galaxy clusters may not help in detection because the noise is
much larger than for small clusters. If a detection is to be made using a 
galaxy survey, it will be important to carry out cross checks. For example,
enough data needs to be a available that clusters can be divided by
mass into more than one bin, to make sure that the effect is largest 
for the high mass clusters (even if the error bar is large). The more
tracers of the clusters potential that are available, the better. Broadhurst
\& Scannapieco (2000) have shown that X-ray emission from intracluster 
gas could be used. Other probes of the potential could include using
redshifts of intracluster planetary nebulae (e.g., Feldmeier \etal 2003).

It may be marginally possible to  detect the gravitational redshift 
of cluster galaxies with the data
available today.
However,  more surveys of greater scope have been imagined for the 
future. Someday this will provide  us enough information
to separate the gravitational redshift effect from total redshift.

We thank the anonymous referee for useful suggestions. RACC 
acknowledges support from the NASA-LTSA program, contract NAG5-11634.

{}

\end{document}